\documentstyle[twoside,fleqn,espcrc2,epsf]{article}

\title{
\vskip -100pt
\mbox{} \hfill BI-TP 2003/26\\
\vskip 80pt
Heavy quark free energies, potentials and the renormalized Polyakov loop\thanks{Presented by O. Kaczmarek. This work is supported by 
BMBF under grant No.06BI102 and DFG under grant FOR 339/2-1. P.P. is Goldhaber
Fellow supported by the contract DE-AC02-98CH10886 with the U.S. Department of Energy.
}}
\author{ O. Kaczmarek\address{Fakult\"{a}t f\"{u}r Physik, Universit\"{a}t 
Bielefeld, D-33615 Bielefeld, Germany},
F. Karsch$^{\rm a}$,
P. Petreczky\address{Physics Department, Brookhaven National Laboratory,
  Upton, NY 11973 USA},
F. Zantow$^{\rm a}$}

\begin{document}

\begin{abstract}
We discuss the renormalized free energy of a heavy quark 
anti-quark pair in the color singlet channel 
for quenched and full QCD at finite temperature. 
The temperature and mass dependence, as well as 
its short distance behavior is analyzed. 
Using the free energies we calculate the heavy quark 
potential and entropy in quenched QCD. 
The asymptotic large distance behavior 
of the free energy is used to 
define the non-perturbatively renormalized Polyakov loop which is 
well behaved in the continuum limit. 
String breaking is studied in the  
color singlet channel in 2-flavor QCD.
\vspace*{-0.4cm}
\end{abstract}
\maketitle
\section{Introduction}
Polyakov loop correlation functions are generally used to analyze the 
temperature dependence of confinement forces and the screening in the
high temperature phase of QCD. The Polyakov loop correlation
functions are directly related to the {\it change in free energy} arising
from the presence of a static quark anti-quark pair in a thermal
medium, $\langle {\rm Tr} L(\vec{x}) {\rm Tr} L^{\dagger}(0) \rangle \sim
\exp(-F_{\bar{q}q}(|x|,T)/T)$ [1,2]. Nonetheless, over the years
it has become customary to call this observable the {\it heavy quark
potential at finite temperature}. The role of any additional entropy
contribution at finite temperature thus has been ignored. We will discuss
here a first calculation of the access energy (potential) from
the access free energy of static heavy quark sources. 

We will
concentrate on results from calculations of the singlet free energy,
$F_1(|x|,T)/T = -\ln \langle {\rm Tr} L(\vec{x}) L^{\dagger}(0) \rangle$
and will present results on color averaged and octet free energies elsewhere.
The operator used to calculate $F_1$ is not gauge invariant.
Calculations thus have been performed in Coulomb gauge. It has been 
shown in [3] that this approach is equivalent to using a suitably
defined gauge invariant (non-local) operator for the singlet free energy.

The Polyakov loop, calculated on the lattice, is ultra-violet divergent
and needs to be renormalized to become a meaningful observable in
the continuum limit. We will do so by renormalizing the free energies
at short distances. Considering that no additional divergences arise
from thermal effects and that at short distances the heavy quark
free energies will not be sensitive to medium effects, renormalization is achieved 
through a matching of free energies to the zero temperature heavy quark 
potential. Using the large distance behavior of the renormalized
free energies we can define the renormalized Polyakov loop which is well
defined also in the continuum limit.  
\vspace*{-1.5mm}
\section{The renormalized free energy}
\vspace*{-1mm}
At small distances the free energy of a quark anti-quark pair should essentially be temperature
\begin{figure}[h]
\vspace*{-13mm}
\centerline{
\epsfxsize=6.7cm\epsfbox{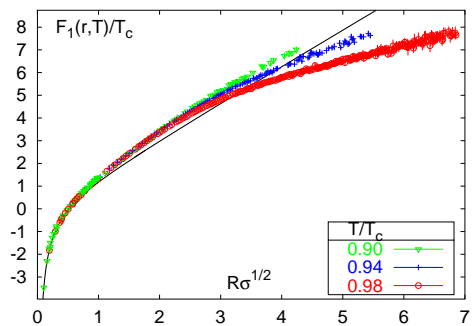}
}
\vspace*{-12mm}
\caption{
Renormalized singlet 
free energy below $T_c$. The solid line indicates the $T$=$0$-potential \cite{Necco:2001xg}.
}
%\vspace*{-8mm}
\label{fig:fesb}
\end{figure}
\newpage
independent at distances which are smaller than 
the average separation between
partons in the thermal medium. Therefore, in the limit of small $R=\vert \vec x \vert$, 
the heavy quark free energy $F_1(R,T)$
is supposed to be given by the heavy quark potential, $V_{\bar q q}(R)$, at zero temperature. 
By matching $F_1(R,T)$ to the $T$=$0$-potential at small $R$, the divergent
self-energy contributions to the free energy are removed. 
\begin{figure}[t]
%\vspace*{-7mm}
\centerline{
\epsfxsize=6.7cm\epsfbox{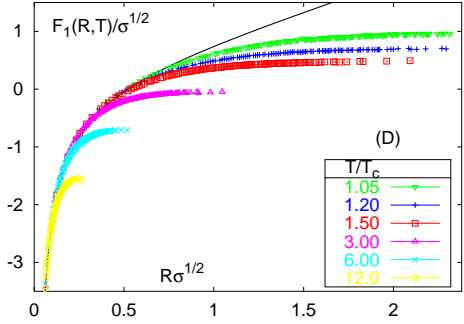}
}
\vspace*{-12mm}
\caption{
Renormalized singlet 
free energy above $T_c$ together with the 
potential at $T$=$0$ \cite{Necco:2001xg}.
}
\vspace*{-8mm}
\label{fig:feso}
\end{figure}

\begin{figure}[b]
\vspace*{-8mm}
\centerline{
\epsfxsize=6.7cm\epsfbox{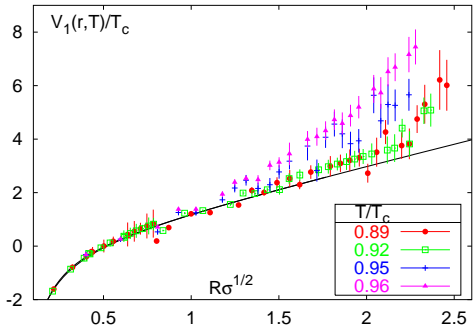}
}
\vspace*{-11mm}
\caption{
Color singlet internal energy at three different values below $T_c$. The solid
line is the zero temperature potential.
}
%\vspace*{-7mm}
\label{fig:csb}
\end{figure}

\begin{figure}[t]
%\vspace*{-7mm}
\centerline{
\epsfxsize=6.7cm\epsfbox{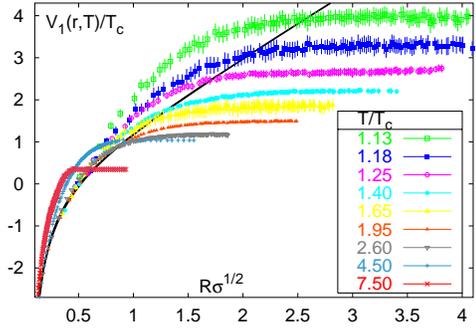}
}
\vspace*{-11mm}
\caption{
Color singlet internal energy above $T_c$.
}
\vspace*{-8mm}
\label{fig:csa}
\end{figure}

The singlet free energy (Fig.~\ref{fig:fesb}) coincides with the $T$=$0$-potential at short
distances and shows no temperature dependence up to $R\sqrt{\sigma}\approx
1$. There is a significant overshooting of $F_1$ relative to the
$T$=$0$-potential which may be understood in terms of
string fluctuations. At large separation $F_1$ shows confining behavior
with a string tension that is decreasing with increasing temperature. This is a
first indication that there are entropy contributions at large separations,
while at small $R$ the entropy is zero. 

Above $T_c$ (Fig.~\ref{fig:feso}) the free energy is again $T$-independent at
small distances and coincides with the $T$=0-potential at sufficiently small
$R$. With increasing temperature screening sets in at shorter separation.

$F_1$ decreases with increasing temperature at fixed $R$, indicating that there
is a positive entropy, $S=-\frac{\partial F_1}{\partial T}$,  contribution at
large distances, while it is close to zero,
due to the (asymptotic) $T$-independence of $F_1$, at small $R$.
\section{Internal energy and entropy}
The free energy has a quite complex $R$-dependence. It is not only
determined by the potential energy but contains entropy contributions, which are
also $R$-dependent ($F_1(R,T) = V_1(R,T) - T S_1(R,T)$). 
The singlet free
energy is temperature independent at sufficiently small distances, while it
gets $T$-dependent at large $R$. This leads to a vanishing entropy at small
separation and a non-vanishing entropy at large $R$.

In order to separate both contributions to the free energy, we calculate the
derivatives with respect to $T$, 
\vspace*{-0.1cm}
\begin{eqnarray}
V_1(R,T) = -T^2 \frac{\partial F_1(R,T)/T}{\partial T} 
%%, S_1 = - \frac{\partial
%%  F_1(R,T)}{\partial T}
\end{eqnarray}
\vspace*{-0.3cm}

In Fig.~\ref{fig:csb} the internal energy for three different temperatures below
$T_c$ is shown. At small distances the data coincide
with the $T$=$0$-potential, while at large separations they tend
toward larger values with increasing temperature. This is in contrast to the
behavior of the free energy (Fig.~\ref{fig:fesb}) and shows that entropy
contributions play a major role at finite temperature.

The internal energies above $T_c$ (Fig.~\ref{fig:csa}) also show major
differences compared to the free energy. In the limit of small separations they
turn into the $T$=$0$-potential. At intermediate $R$ 
the potential becomes narrower (overshooting) which is a consequence of
screening and finally leads to a flat potential
at large
distances. This behavior indicates that also above $T_c$ the free energy is
not only controlled by the internal energy itself.
\vspace*{-0.1cm}
\section{The renormalized Polyakov loop}
Using the renormalized free energies
from Fig.~\ref{fig:feso} we can define the renormalized Polyakov loop
\cite{Kaczmarek:2002mc},
\begin{eqnarray}
L^{\mathrm ren} = \exp\left( - \frac{F_1(R=\infty,T)}{2 T}\right).
\label{renpo}
\end{eqnarray}
In quenched QCD it is zero below $T_c$ by construction. $L^{\mathrm ren}$ does not depend on
$N_\tau$ and is well behaved in the continuum limit.

\begin{figure}[htbp]
\vspace*{-7mm}
\centerline{
\hspace*{-0.5cm}\epsfxsize=6.7cm\epsfbox{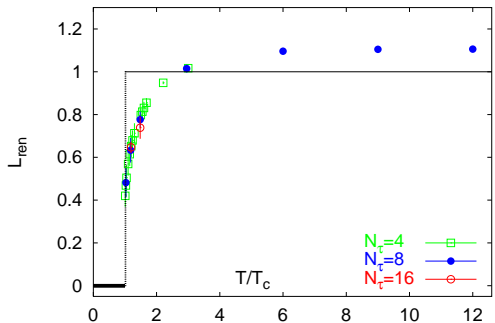}
}
\vspace*{-12.5mm}
\caption{
The renormalized Polyakov loop, $L^{\mathrm ren}$.
}
\vspace*{-10mm}
\label{fig:renp}
\end{figure}
\vspace*{-0.1cm}
\section{Full QCD}
\begin{figure}[t]
\vspace*{-1.5mm}
\centerline{
\epsfxsize=6.7cm\epsfbox{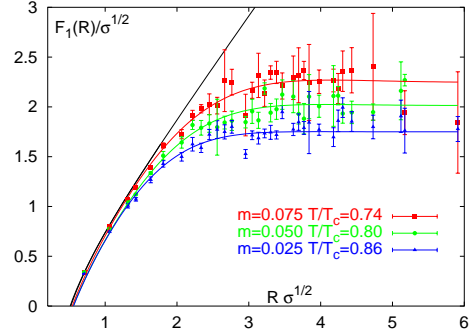}
}
\vspace*{-11.5mm}
\caption{
Singlet free energies for 2-flavor full QCD at $\beta=3.48$ on $16^3\cdot 4$
lattices.
}
\vspace*{-8mm}
\label{fig:giac}
\end{figure}

The concepts presented above can be extended to the case of full QCD, 
where string breaking has been observed in Polyakov loop correlation functions
in \cite{DeTar:1998qa}.
In Fig.~\ref{fig:giac} the renormalized singlet free energies in 2-flavor QCD for
three different quark masses are shown. They tend towards the $T$=$0$-potential 
at small $R$ and get constant at large distances due to the breaking
of the string. 
This behavior might be explained by the model for string breaking proposed in
\cite{giacomo},
\begin{eqnarray}
\vspace*{-0.1cm}
F_1(R) = \left(\sigma R - \frac{\pi}{12 R}\right) e^{-\omega R^2} - \left(1 -
  e^{-\omega R^2}\right) A
\label{giacf}
\vspace*{-0.1cm}
\end{eqnarray}
The lines in Fig.~\ref{fig:giac} show the results of a fit of (\ref{giacf})
using the zero temperature string tension, $\sigma$, and two fit parameters
$\omega$ and $A$. The data seem to be well described by this Ansatz. 

\vspace*{-0.15cm}
\section{Conclusions}
\vspace*{-0.1cm}
We have discussed the renormalized color singlet free energy in quenched QCD as well as
2-flavor QCD. 
It was shown that $F_1$ is not only given by the potential energy, but entropy
contributions play a major role at large distances. 
First results for the potential energy were presented and show a quite different
behavior compared to the free energy. 

The results in 2-flavor QCD show that the concepts, developed in quenched QCD,
can be extended to full QCD. 
The data can be well described by a string model inspired Ansatz proposed in
\cite{giacomo}.
A more detailed analysis of the temperature and mass dependence, as well as a
closer look to the short distance regime, is needed.

The large distance behavior of the free energy was used to calculate the
renormalized Polyakov loop, which is well behaved in the continuum limit.


\vspace*{-0.2cm}
\begin{thebibliography}{99}
\vspace*{-0.1cm}

%\cite{Kaczmarek:1999mm}
\bibitem{Kaczmarek:1999mm}
L.~G.~McLerran and B.~Svetitsky,
Phys.\ Rev.\ D {\bf 24} (1981) 450
%[arXiv:hep-lat/9908010].
%%CITATION = HEP-LAT 9908010;%%

%\cite{Kaczmarek:2002mc}
\bibitem{Kaczmarek:2002mc}
O.~Kaczmarek {\it et al.},
%``Heavy quark anti-quark free energy and the renormalized Polyakov loop,''
Phys.\ Lett.\ B {\bf 543} (2002) 41
%%CITATION = HEP-LAT 0207002;%%

%\cite{Philipsen:2002az}
\bibitem{Philipsen:2002az}
O.~Philipsen,
%``Non-perturbative formulation of the static color octet potential,''
Phys.\ Lett.\ B {\bf 535} (2002) 138
%[arXiv:hep-lat/0203018].
%%CITATION = HEP-LAT 0203018;%%

%\cite{Necco:2001xg}
\bibitem{Necco:2001xg}
S.~Necco and R.~Sommer,
%``The N(f) = 0 heavy quark potential from short to intermediate  distances,''
Nucl.\ Phys.\ B {\bf 622} (2002) 328
%[arXiv:hep-lat/0108008].
%%CITATION = HEP-LAT 0108008;%%

%%\cite{Zantowxx}
%\bibitem{Zantowxx}
%O.~Kaczmarek {\it et~al.}, Phys.\ Lett.\ B {\bf 543} (2002) 41

%\cite{DeTar:1998qa}
\bibitem{DeTar:1998qa}
C.~DeTar {\it et~al.},
Phys.\ Rev.\ D {\bf 59} (1999) 031501
%%CITATION = HEP-LAT 9808028;%%

\bibitem{giacomo}
D.~Antonov {\it et al.},
JHEP 0308 (2003) 011
%[arXiv:hep-lat/9908010].
%%CITATION = HEP-LAT 9908010;%%


\end{thebibliography}
\end{document}